\documentclass[twoside,slac_two]{revtex4}
\usepackage{graphicx}
\usepackage{fancyhdr}
\newcommand{\be}{\begin{equation}}
\newcommand{\ee}{\end{equation}}
\newcommand{\bea}{\begin{eqnarray}}
\newcommand{\eea}{\end{eqnarray}}
\newcommand{\bean}{\begin{eqnarray*}}
\newcommand{\eean}{\end{eqnarray*}}
\newcommand{\qq}{$q\bar{q}$}
\newcommand{\cc}{$c\bar{c}$}
\newcommand{\ssbar}{$s\bar{s}$}
\newcommand{\bb}{$b\bar{b}$}

\newcommand{\QQ}{$Q\bar{Q}$}
\begin{document}

\title{Three flavours of Hybrid or $\pi$ exhange: which is more attractive? }

%

\author{F.E.Close}
\affiliation{Rudolf Peierls Centre for Theoretical Physics;
University of Oxford; Oxford OX1 3NP; England}

\begin{abstract}
This review summarises issues that have arisen since the appearance of ``Rumsfeld Hadrons".
We show that signals $\phi(2175)$; $Y(4260)$ and $\Upsilon(10890)$ in the \ssbar~,\cc~ and \bb~
share features that point to the possible role of $\pi$ exchange forces between flavoured mesons
generating effects
that can mimic hybrid mesons. The flavour dependence of these phenomena may help to 
resolve this question.

\end{abstract}

\maketitle

\thispagestyle{fancy}


\section{Introduction}

The first part of my talk replicated much of what was reported in ``Rumsfeld Hadrons"
and will not be repeated here. For that see\cite{fecbled}. Here I review some subsequent developments 
concerning the possible discovery of hybrid charmonium. In summary: there are undoubtedly signals in 
the $1^{--}$ wave that have the a priori
character of hybrid mesons. The data in the charm sector seem to rule out a particular tetraquark 
interpretation and are consistent with a charmonium hybrid. However, there are reasons to consider the 
role of $\pi$
exchange, which gives an attraction in this channel. Comparisons of hybrid predictions and 
attractive forces from $\pi$ exchange as a function of flavour may resolve this question.

 \section{Charmonium: the $Y(4260)$}

First let's consider the lightest of the novel charmonium states, the $X(3872)$ at $D^0D^{*0}$
threshold.

 This state now appears to have
$C=+$ and be consistent with $1^{++}$~\cite{3872data}. This $J^{PC}$
was first suggested in Ref.~\cite{fcpage3872} and a dynamical picture of it
as a quasi-molecular $D^{\ast 0}\bar{D}^0$ state discussed
in Refs.~\cite{fcpage3872,swanson3872}.
In $e^+e^- \to \psi + X$ there is no sign of the $X(3872)$: the suppression of this state among
prominent $C=+$ charmonium states~\cite{belleddstar} is 
consistent with its molecular versus simple \cc~ nature.

It is generally agreed that the $X(3872)$ has a 
tetraquark affinity; whether it is a genuine $D^0D^{*0}$ molecule or
a compact $cu\bar{cu}$ is a more subtle issue. If the quark-pairs are tightly clustered into 
di-quarks, then a $S=0$
and $S=1$ are required to make the $1^{++}$. Consequently other states, combinations of $0^+-0^+$ and 
$1^+-1^+$
would be expected. The absence of such a rich spectrum suggests that the overriding dynamics is that 
the constituents
rearrange into loosely bound colour singlet $c\bar{u}$-$u\bar{c}$, or $D^0D^{*0}$.

That such a molecular state can be generated by the attractive force of $\pi$ exchange
was suggested in 
ref\cite{cp04,swanson3872}. 
Ref\cite{oset} also has discussed the dynamical generation of such states.
What is particularly interesting is that it had been predicted long ago that the $\pi$ exchange
that is known to bind the deuteron, may also act between metastable mesons and cause attractions in
certain channels\cite{torn}, which Tornqvist referred to as ``deusons". One such example was the 
$D\bar{D^*}$
channel. A test of this picture is that
$\pi$ exchange can also occur between $D$ and $D^*$ (i.e with no $\bar{D}$ \cite{torn})
leading to structures in channels with charm = $\pm 2$. Searching for such states among the debris at 
KEK, GSI and LHC
could be important in isolating evidence for this $\pi$ exchange dynamics. 

Before invoking exotic explanations of the various signals that have recently appeared in charmonium,
such as $Y(4260)$ in $\psi \pi\pi$, $X(4350)$ in $\psi'\pi\pi$ and $X(4430)$ in $\psi'\pi$, we should 
satsify ourselves
that there are not more mundane explanations. As each of these states is near an S-wave threshold 
involving
charmed mesons that are metastable on the timescale  of the strong interaction, then the role of $\pi$ 
exchange
here needs to be assessed.

\section{Attractive $\pi$ exchange}

The analysis that Tornqvist applied to $D\bar{D^*}$ can be applied to other combinations of $D,D^*, 
D_0,D_1$ and their
charge conjugates, and to their bottom analogues. I have been looking at this in collaboration with 
Qiang Zhao and Christopher Thomas
\cite{ctz}. This is relevant as the three novel states are in the vicinity of the S-wave thresholds
$D^*D_0$ and $DD_{1L}$ ($Y(4260)$); $DD_{1H}$ and $D^*D_2$ ($X(4350)$); $D^*D_1$ ($X(4430)$).

The basic idea is that $\pi$ exchange has both a direct Yukawa term $C(r)$ and also a tensor 
interaction $T(r)$
that links $S-D$ waves\cite{torn}. In the deuteron there is attraction within the $^3S_1$; repulsion 
in $^3D_1$ but an attractive coupling between these waves that enables binding. The effective 
potential in a basis of $^3S_1$ and $^3D_1$ states may be summarised as  

\begin{equation}
V= -\frac{25}{3}V_0\big[ 
\left( \begin{array}{cc}
1 &  0\\
0 & 1
\end{array}\right)C(r)
+
\left( \begin{array}{cc}
0 &  \sqrt{8}\\
\sqrt{8} & -2 
\end{array}\right)T(r)\big].
\label{eqd}
\end{equation}
The known binding of the deuteron normalises the strength of the above, which Tornqvist\cite{torn} 
then applied  to mesons, for example the $D\bar{D^*} \pm D^*\bar{D}$
case (note the two charge conjugation eigenstates which will have opposite overall signs, leading to
attraction in one and repulsion in the other). 
For the $1^{++}$ channel this becomes\cite{torn}

\begin{equation}
V= -3V_0\big[ 
\left( \begin{array}{cc}
1 &  0\\
0 & 1
\end{array}\right)C(r)
+
\left( \begin{array}{cc}
0 &  -\sqrt{2}\\
-\sqrt{2} & 1 
\end{array}\right)T(r)].
\label{eqdd}
\end{equation}
Notice that relative to the deuteron, in this case the signs in the tensor interaction are inverted 
relative to the deuteron and binding is
enhanced.

The flavour dependence of the binding is interesting in that \cite{torn} noted that
the net attraction is greater for heavy flavours such as  than for light;
$B\bar{B^*}$ being bound; $D\bar{D^*}$ being around threshold and $K\bar{K^*}$ being an attractive
enhancement above threshold. At first sight this appears a paradox as for heavy-light states, it is 
only the light flavour that couple to the pion, and the heavy
flavour is apparently a passive spectator, so how can its mass affect the result? This is because the 
larger
kinetic energy of the light flavoured states tends to counterbalance the overall potential leading 
tobetter
binding of bottom than charm, and of charm than strange. Hence Tornqvist gives only tentative
suggestions about possible attractions in the strange sector; for the charm sector there begins to be 
hints of possible states appearing at threshold, which may now have some confirmation in the case of 
the $1^+(3872)$; for bottom mesons the model appears to imply some bound states should occur.

This pattern will have relevance later when we consider the flavour dependence of hybrid meson masses.

We find that the structure of eq(\ref{eqdd}) also applies in the cases of $DD_1$ and $D^*D_0$ in
$1^{-}$.
Note that parity conservation requires the $\pi$ vertex to link $D \leftrightarrow D^*$; and $D_0 
\leftrightarrow D_1$.
In the heavy quark limit the latter is $D_{1L}$; in general $\pi$-exchange couples 
the $D_0$ to the $p_{1/2}$ combination of each of the two axial mesons.
These remarks apply also to the charge conjugate states. Consider first the S-wave $1^-$ channel 
accessible to
$D_1\bar{D}$ and $D_0\bar{D^*}$ (or their charge conjugate analogues, which is understood always). The 
$\pi$ exchange gives an off diagonal
potential linking $D_1\bar{D} \leftrightarrow D_0\bar{D^*}$.

In the case of $D\bar{D^*} \pm D^*\bar{D}$ Tornqvist found attraction in I=0 $1^{++}$ and repulsion in 
$1^{+-}$. 
In the off-diagonal case, $D_1\bar{D} \leftrightarrow D_0\bar{D^*}$ $(+ c.c.)$
we find that the channels for strong attraction are isoscalar for {\it both} $1^{--}$ and $1^{-+}$.
Thus we find attractions in the $1^{--}$ channel, where the $Y(4260)$ and possibly $Y(4350)$ are seen 
coupling to $e^+e^-$,
and also in the exotic $1^{-+}$ channel. It is intriguing that it is in these $1^{-\pm}$ channels 
and in this mass region where hybrid mesons are also predicted to occur; furthermore they
are predicted to have 
preferred couplings to these very $DD_1$ and $D^*D_0$ states. Possible implications of this will be 
discussed later.

In the case of $Z(4430)$ seen in $\psi' \pi$, Bugg\cite{bugg} has also noticed the nearness to the 
$D^*D_1$ threshold
and argued that the existence of this state and the threshold could be linked. In the absence of
a model for the attractive force he was unable to predict the $J^P$, but if we apply the $\pi$ 
exchange analysis to this case (which is the last
remaining combination of thresholds for S and P wave $c\bar{q}$ states) then we can predict possible 
quantum numbers.
Here we will consider first the $D^*\bar{D^*}$ and the $^1S_0$ and $^5D_0$ basis for which

\begin{equation}
V= -\frac{\gamma}{2} V_0\big[ 
\left( \begin{array}{cc}
2 &  0\\
0 & -1
\end{array}\right)C(r)
+
\left( \begin{array}{cc}
0 &  \sqrt{2}\\
\sqrt{2} & 2 
\end{array}\right)T(r)].
\label{eqdd}
\end{equation}
where the overall scale factor $\gamma$ depends on the $J^P$ and is discussed below. The same pattern 
emerges for
the $D^*\bar{D_1}$ (and charge conjugate).

The tensor term adds to the attraction
if the leading term is already attractive. For other spin combinations the matrix structure is more 
involved but
the tensor terms do not essentially alter the attraction or repulsion systematics of the leading term. 
Thus
in S-wave the couplings are to $(0,1,2)^-$ and one finds the following attractive channels,
which are listed in the sequence $J^P$, isospin, relative strength of attraction.

In the heavy quark limit, at $D^*D_{1L}$ we have for the scale factors $\gamma$

\bea
0^- ~~I=0 ~~ 12 \nonumber \\
1^- ~~~ I=0 ~~~ 6 \nonumber \\
2^- ~~~I=1 ~~~ 2
\eea
while at $D^*D_{1H}$ we find attraction in the complementary channels
\bea
0^- ~~~I=1 ~~~ 2 \nonumber \\
1^-~~~ I=1 ~~~ 1 \nonumber \\
2^- ~~~I=0 ~~~ 3
\eea

If this state with both hidden charm and isospin is confirmed then it definitely goes beyond 
charmonium and demands
tetraquarks. To distinguish $\pi$ exchange molecules from tight clustered diquark-antidiquark will 
involve finding other
examples and collating their $J^P$ pattern. The pattern from $\pi$ exchange differs from the richer 
spectroscopy of
tetraquarks. In particular the most likely $J^P$ would be $0^-$ or $2^-$. However, one would expect
even larger effects in the $I=0$ sector in the $0^-$ and also $1^-$. Thus if this
signal is driven by attractive $\pi$-exchange, then some signal in this same
mass region should also occur in $e^+e^- \to \psi \eta/\eta'$ with at least as big strength.

A challenge for dynamics is also to explain why the state is seen in $\pi \psi'$ but not apparently in 
$\pi\psi$.
As conjectured by Bugg\cite{bugg}, it is possible to force a suppression due to nodes in 
wavefunctions\cite{thomas} 
 but it is
not an overwhelming effect and while such a state should be expected to have some strength in $\psi 
\pi$, one should
also
anticipate I=0 partners in $\psi \eta$ among other channels.

A final reminder\cite{torn}: $\pi$ exchange also leads to potential bound states in double charm 
combinations,
not just in the hidden charmonium channel. In principle this could discriminate these molecular
combinations from others, such as hybrid charmonium, though their associated production in order
to conserve charm will imply many particle final states and impose severe challenges to analysis.

I will now focus on the best established state, the $Y(4260)$\cite{babar}, and review interpretations 
as hybrid \cc~\cite{cp05,pene} or $cs\bar{c}\bar{s}$
tetraquark\cite{maiani}. The data already appear to disfavour the latter. The challenge will be to 
distinguish the former from
the $\pi$ exchange. One way will be to look for analogues in the \ssbar~ and \bb~ sectors.

\section{Hybrid Quarkonium: Theory}

Mass predictions for the $J^{PC}$ exotic $1^{-+}$ hybrids were reviewed in ref\cite{bcs}.
Previous results based on lattice QCD, such as flux tube models, had assumed an adiabatic 
approximation.
Ref\cite{bcs} made numerical studies that relaxed that assumption and found that hybrid signals
should arise in the following mass regions

\bea
s\bar{s}: ~~2.1-~2.2 GeV \nonumber \\
c\bar{c}: ~~4.1-~4.2 GeV \nonumber \\
b\bar{b}: 10.8-11.1 GeV
\eea
The adiabatic approximation was found to be good for \bb~ and reasonable for the lighter flavours.

In view of the signals that could be candidates for the $1^{--}$ hybrids, which we shall discuss 
later, let's first look at these in more detail.

Eight low-lying hybrid charmonium states were
predicted in the flux-tube model to occur at $4.1-4.2$ GeV~\cite{bcs},
and in UKQCD's quenched lattice QCD calculation with infinitely heavy
quarks the exotic $1^{-+}$ was predicted to be $4.04\pm 0.03$ GeV (with un-quenching estimated to 
raise
the mass by $0.15$ GeV)~\cite{ukqcd}. 
Quenched lattice QCD indicates that the $c\bar{c}g$
$1^{--},\; (0,1,2)^{-+}$ are less massive than
$1^{++},\; (0,1,2)^{+-}$~\cite{juge}. The spin splitting for this lower
set of hybrids in quenched lattice NRQCD is
$0^{-+}<1^{-+}<1^{--}<2^{-+}$~\cite{drum}, at least for $b\bar{b}g$.
This agrees with the ordering found in the model-dependent
calculations for $q\bar{q}g$~\cite{bcd} in the specific case of
$c\bar{c}g$~\cite{pthesis,merlin} though there is considerable uncertainty
in the magnitudes\cite{drum}. In particular, the cavity QCD calculations have not included
the contribution from four-gluon vertex; although higher order in $\alpha_s$. it is 
possible such contributions are not negligible\cite{ls}.

Thus the consensus is that the resulting pattern
is, in decreasing mass, $1^{--}; 1^{-+}; 0^{-+}$ with the mass gap between each state being 
the same and of the order
of 10-100MeV. Thus theory strongly
indicates that if $Y(4260)$ is $c\bar{c}g$, and the splittings are not
due to mixing or coupled channel effects, then the $J^{PC}$ exotic
$1^{-+}$ and non-exotic $0^{-+}$ $c\bar{c}g$ are below
$D^{\ast\ast}\bar{D}$ threshold, making them narrow by virtue of the
selection rules. The $1^{-+}$ decay modes~\cite{dunietz} and branching
ratios~\cite{cgodfrey} have extensively been discussed. Thus on the basis of masses alone,
it is consistent to identify possible states as 
$1^{--}(4.25); 1^{-+}(4.1);0^{-+}(3.9)$ and to speculate whether there are two states
$1^{-+}(4.1);0^{-+}(3.9)$ in either the $X/Y(3940)$ structures of Belle or $e^+e^-\to \psi + X$. 
This is clearly a question that statistics from a super-B factory may resolve for the
B-decays or $e^+e^- \to \psi + X$.

To the extent that these spin dependent arguments are relevant to the more realistic situation, one 
expects for hybrid \bb~
state that the mass of the $1^{--}$ state will be similar or at most some tens 
of MeV more massive than the $1^{-+}$, whereas for \ssbar~ the splitting could be 
$O(100)$MeV\cite{bcd}. In practice
I suspect that the strong S-wave coupling of such states to flavoured channels that are near to 
threshold, such as \cc~ $\to DD_1; D^*D_0$ in the case of charmonium, may cause significant
mass shifts\cite{bs} and potentially dominate the spin-dependence of such masses.

Ref\cite{cp95} identified prominent hadron decays modes for such masses to include
\ssbar $\to KK_1(1400;1270)$ in S-wave and $KK_2$ in D-wave; . The
analogous situation for \bb~ would be to $BB_1$ and $B^*B_0$. Given that
the anticipated hybrid masses are already in the vicinity of thresholds to which they are predicted 
to have strong S-wave couplings, and further the fact that
$\pi$ exchange is predicted to give attractions among these mesons in the overall I=0 $1^{-\pm}$ 
channels, it would be
surprising if signals were not seen in these $1^{-\pm}$ modes at least. Determining whether
they are pure molecule or require a short range \QQ~ seed, is one challenge; if such a seed
is present, then for the $1^{--}$ case we would need to determine whether it has S=1 (as for a 
conventinoal quarkonium) 
or S=0 (as for a hybrid). To do so will require studies of \cc~, \bb~ and \ssbar.

\section{Hybrid Quarkonium: Phenomena}

Mass arguments alone will not be convincing; we need to understand the dynamics of production and
decay and show that these fit best with hybrid states.
I now turn to hybrid charmonium and evaluate the prospects that it is being exposed in the
enigmatic vector state $Y(4260)$ which is seen in $e^+e^- \to \psi \pi\pi$, with no observed decay 
into $D\bar{D}$. The mass, large width into $\psi\pi\pi$, small leptonic width ($O(5-80)$eV, contrast
$O$(keV) for known states), affinity for $DD_1$ threshold and
apparent decay into $\psi \sigma$ or $\psi f_0(980)$ are all consistent with predictions made for
hybrid vector charmonium\cite{cp05}.

The fact that there is no sign of established
3S/2D(4040/4160)
4S(4400)
in the $\psi\pi\pi$ data already marks this state as anomalous, and 
its characteristics are tantalisingly
similar to what has been predicted for hybrid charmonium. However, the fact that
it is near the $DD_1$ threshold might be the reason for the large $\psi\pi\pi$ signal
independent of its nature:
the $DD_1$ are produced in S-wave, with small relative momenta and as such there is every
likelihood that they can interchange constituents, leading to \cc~ + \qq~ final states, such as $\psi 
\pi\pi$, without any $O(\alpha_s)$ suppressions from intermediate perturbative gluons (in contrast
to the case for $\psi(3685)$ and \cc~ resonances at other masses).

A lattice-inspired flux-tube model showed that the decays of hybrid
mesons, at least with exotic $J^{PC}$, are suppressed to pairs of
ground state conventional mesons~\cite{ipaton,ikp}.  This was extended
to all $J^{PC}$, for light or heavy flavours in Ref.~\cite{cp95}.
A similar selection rule was found in constituent gluon
models~\cite{pene}, and
their common quark model origin is now understood~\cite{pagesel}.
It was further shown that these selection rules for light flavoured
hybrids are only approximate, but that they become very strong for
$c\bar{c}$~\cite{cp95,pthesis}.  This implied that decays into
$D\bar{D},\: D_s\bar{D}_s,\: D^\ast \bar{D}^\ast$ and $D_s^\ast
\bar{D}_s^\ast$ are essentially zero while $D^\ast\bar{D}$ and
$D_s^\ast\bar{D}_s$ are very small, and that $D^{\ast\ast}\bar{D}$, if above
threshold, would dominate. (P-wave charmonia are
denoted by $D^{\ast\ast}$).  As $c\bar{c}g$ is predicted around the
vicinity of $D^{\ast\ast}\bar{D}$ threshold, the opportunity for
anomalous branching ratios in these different classes was proposed as
a sharp signature~\cite{cp95,bcs}.

 More recently the signatures for hybrid charmonia were expanded to
note the critical region around $D^{\ast\ast}\bar{D}$ threshold as a
divide between narrow states with sizable branching ratio into
$c\bar{c}\; +$ light hadrons and those above where the anomalous
branching ratios would be the characteristic
feature~\cite{dunietz,cgodfrey}.  It
was suggested to look in $e^+e^-$ annihilation in the region
immediately above charm threshold for state(s) showing such anomalous
branching ratios~\cite{cgodfrey}. The leptonic couplings
 to $e^+e^-,\; \mu^+\mu^-$ and $\tau^+\tau^-$  were expected
to be suppressed~\cite{ono}
(smaller than radial S-wave $c\bar{c}$ but larger
than D-wave $c\bar{c}$, but with some inhibition due to the fact that
in hybrid vector mesons spins are coupled to the $S=0$, whose coupling
to the photon is disfavoured~\cite{cgodfrey}).

Thus several of the theoretical expectations for
$c\bar{c}g$ are born out by $Y(4260)$: (1) Its mass is
tantalizingly close to the prediction for the lightest hybrid
charmonia; (2) The expectation that the $e^+e^-$ width should be
smaller than for S-wave $c\bar{c}$ is consistent with
the data\cite{cp05};
(3) The predicted affinity of hybrids to $D^{\ast\ast}\bar{D}$
could be related to the appearance of the state near the
$D^{\ast\ast}\bar{D}$ threshold. The formation of $D^{\ast\ast}\bar{D}$
at rest may lead to significant re-scattering into $\psi\pi^+\pi^-$,
which would feed the large signal;
(4) The absence of any enhancement in 
``ground state charm" such as $D\bar{D}, D^*\bar{D}, D^*\bar{D^*}, D_s\bar{D_s}$ etc
is also an explicit signature for hybrid charmonium.

It has become increasingly clear recently that there is an affinity
for states that couple in S-wave to hadrons, to be attracted to the
threshold for such channels~\cite{hadron03}. The hybrid candidate
$1^{--}$ appearing at the S-wave $D_1(2420)\bar{D}$ is thus interesting.
However, one could argue that {\it any} \cc~ resonance in this
region would be attracted likewise, so these phenomena do not necessarily imply a hybrid meson
rather than a conventional \cc~ as the source. Ways of distinguishing these are discussed later.

The nearness of $Y(4260)$ to the $D_1(2420) \bar{D}$ threshold, and
to the $D_1' \bar{D}$ threshold, with the broad $D_1'$ found at a mass
of $\sim 2427$ MeV and width $\sim 384$ MeV~\cite{dmass}, indicate
that these states are formed at rest. Also, these are the lowest open
charm thresholds that can couple to $1^{--}$ in S-wave (together with
$D_0 \bar{D}^\ast$, where the $D_0$ mass $\sim 2308$ MeV and
width $\sim 276$ MeV~\cite{dmass}).  Flux-tube model predictions are
that the D-wave couplings of $1^{--}\; c\bar{c} g$ to the $1^{+}$ and
$2^{+}$ $D^{\ast\ast}$ are small~\cite{cp95,pthesis,pss}; and there is
disagreement between various versions of the model on whether the
S-wave couplings to the two $1^{+}$ states are large. If these
couplings are in fact substantial, the nearness of $Y(4260)$ to the
thresholds may not be coincidental, because coupled channel effects
could shift the mass of the states nearer to a threshold that it
strongly couples to; and it would experience a corresponding
enhancement in its wave function. The broadness of $Y(4260)$ also
implies that its decay to $D_1(2420) \bar{D},\; D_1' \bar{D}$ and
$D_0(2308)\bar{D}^\ast$ which feed down to $D^\ast \bar{D} \pi$ and $D
\bar{D} \pi$~\cite{cs05} would be allowed by phase space and
should be searched for to ascertain a significant coupling to
$D^{\ast\ast}$.

Flux-tube model width predictions for other charm modes
are $1-8$ MeV for $D^\ast \bar{D}$~\cite{pss},
with $D\bar{D},\: D_s\bar{D}_s,\:
D^\ast \bar{D}^\ast$ and $D_s^\ast \bar{D}_s^\ast$ even more suppressed.
 Thus a small $D\bar{D}$ and $D_s\bar{D}_s$ mode could single out the hybrid
interpretation, which is very different from the $c\bar{s}s\bar{c}$
four-quark interpretation for $Y(4260)$ which decays predominantly in
$D_s \bar{D}_s$~\cite{maiani}. 

The data\cite{eecharm} on $e^+e^- \to D_s \bar{D_s}$ show
a peaking above threshold around 4 GeV but no evidence of affinity for a structure at 4.26GeV.
This is suggestive and if these data are confirmed, then as well as ruling out a $cs\bar{cs}$ at this 
mass, they will also
add support to the hybrid interpretation. 
The same data also show there is no significant coupling of $Y(4260)$ to $D\bar{D}; D^*\bar{D}$ or 
$D^*\bar{D^*}$, 
all of which are in accord with predictions for a hybrid state.

\subsection*{Theory \ssbar, \cc~ and \bb~ }

If the large $\psi \pi\pi$ signal is solely due to the presence of S-wave $DD_1;D^*D_0$ thresholds and
constituent interchange, there should be analogous phenomena in $\phi \pi\pi$ and $\Upsilon \pi\pi$
associated with the corresponding flavoured thresholds. By seeking evidence for these channels, and 
comparing
any signals, or lack of, it may be possible to identify the dominant dynamics. 

For example, if a $1^{--}$ hybrid meson of \ssbar, \cc~ or \bb~ flavours is involved, and the lattice 
QCD or
flux-tube models are reliable guide to the masses, then we anticipate activity in the energy regions

\bea
s\bar{s}: ~~2.1-~2.2 GeV \nonumber \\
c\bar{c}: ~~4.1-~4.2 GeV \nonumber \\
b\bar{b}: 10.8-11.1 GeV
\label{masses}
\eea
If the effect is simply due to S-wave threshold, without any direct channel resonant enhancement, then 
for \bb~
we need to look in the vicinity of $BB_1$ and $B^*B_0$ thresholds. The mass splitting of $B$ and $B^*$
is 46 MeV; that of the $B_2$ and $B_{1H}$ is 26 MeV. The threshold for $BB_{1H}$ is 11.00GeV; we
expect that $\pi$-exchange effects arise near to the threshold for $BB_{1L}$ and $B^*B_0$
and so we need to estimate at what energy this is. In the case of $c\bar{d}$ the $D_{1L}$-$D_{1H}$
mass gap is some 60 MeV in theory but the data could have them nearly degenerate. The $D_s$ sector is 
confused by the
light $D_{s0}$ and $D_{s1}$, which may be quasi-molecular. A similar possiblity cannot be ruled out in 
the
$B$ sector. Thus what at first sight appeared to be a straightforward question in the
$B$ sector, namely where are the thresholds and will $\pi$ exchange create attractions, is less
clear. We may anticipate that the thresholds for $BB_1$ and $B^*B_0$ are certainly below 11.00 GeV,
perhaps by as much as 200 MeV, and that the splitting between them will be of the order of tens of 
MeV. Thus, here again we find the S-wave threshold region and the prediction for vector hybrid to be 
very similar.

For \ssbar~ we have

\bea
KK_1(1270) = 1760 ~MeV \nonumber \\
KK_1(1410) = 1900 ~MeV \nonumber \\
K^*K_0 ~~  = 2320 ~MeV.
\eea
though the latter pair at least are smeared over some 200 MeV due to widths. The $\pi$-exchange 
attractions
are expected to occur above threshold in the \ssbar~ sector (see remartks in section III), and once 
again,
in the vicinity of the predicted mass of the hybrid $1^{--}$.

Thus in all flavours, we expect vector hybrids, coupling strongly to $0^-1^+$ or to $1^-0^+$
in S-wave, to be amplified by the $\pi$-exchange and to be manifested around these thresholds.
Consequently we should expect signals (perhaps being wise after the event!) and the challenge is to
determine if they are resonant, and if so, whether they are hybrid or conventional.

The spread in threshold masses for \ssbar~ and the expected near-degeneracy for \bb~ would become 
exact degeneracy in the heavy quark
limit $M_{s,c,b} \to \infty$. This has a consequence for the nature of the $\pi\pi$+ [\ssbar, \cc~, 
\bb~] final state.
If there is a direct channel $1^{--}$ \ssbar, \cc~ or \bb~ resonance feeding S-wave flavoured mesons, 
which rearrange
their constituents to give the superficially OZI violating $\pi\pi$+ [\ssbar, \cc~, \bb~] final state, 
then the spin of
the heavy flavours is preserved in the $M_Q \to \infty$ limit. In that limit, a hybrid \cc~ vector 
meson (whose \cc~
are coupled to zero!) would dominantly feed
channels where \cc~ has spin zero; hence $h_c \eta$ rather than $\psi \pi\pi$\cite{maianipriv}. 
Conversely, a conventional $\psi^*$, \cc~
coupled to spin one, would naturally feed the $\psi \pi\pi$. Thus the ratios of branching ratios to 
\cc~ (S=0 or 1) and
light hadrons can test the nature of the initial spin state and distinguish hybrid from either 
conventional resonance or
$\pi$-exchange.

When the meson loops are calculated, there is a destructive interference between the $DD_1$ and 
$D^*D_0$ channels.
This is exact in the degenerate case, which applies in the $M_Q \to \infty$ limit. Conversely, for 
non-degenerate
channels the cancellation fails. Thus in the case of \ssbar~ there is little to be learned, while for 
\bb~ it should
be a clean test; for \cc~ it is indeterminate until such time as $h_c$ or $\eta_c$ channels are 
quantified.

A further strategic test is to measure the polarization of the respective vector mesons in $\pi\pi$ + 
$\phi; \psi; \Upsilon$.
Predicting it is highly model dependent but a similar, or monotonic behaviour with flavour, would hint 
at a common origin
whereas significantly different amounts of polarization could reveal more than one dynamics is 
important.

\subsection*{Phenomena in \ssbar, \cc~ and \bb~ }

Intriguing phenomena are showing up not just in the \cc~, but also in both the \ssbar~ and \bb~ 
sectors. 
 
The cross section for $e^+e^- \to K^+K^-\pi^+\pi^-$ has significant contribution from $e^+e^- \to 
KK_1$
with rescattering into $\phi\pi\pi$. A resonance with width $\Gamma = 58 \pm 16 \pm 20$MeV with
large branching ratio into $\phi \pi\pi$ is seen with mass of 2175MeV\cite{2175,ss}. Not only does the 
mass agree
with the predictions of eq\ref{masses},
simple arithmetic shows that the mass gap from this state to 
$m(\phi)$ is within the errors identical to that between $Y(4260)$ and $m(\psi)$. 
This coincidence (?) has also been noticed by Jon Rosner\cite{rosnermoriond}. This is perhaps
reasonable if the cost of exciting the gluonic flux-tube is not sensitive to the masses of the
\qq~ involved (as lattice QCD seems to suggest), in which case a hybrid vector production and decay
is consistent with data. The $KK_1$ and $K^*K_0$ thresholds do not relate so readily to the 2175 state
as do the analogous charm states with the 4260, which makes it less likely perhaps that the 4260 and 
2175 can be
simply dismissed as non-resonant effects associated solely with S-wave channels opening. 

During this conference evidence for similar happenings in the \bb~ sector\cite{george}. The decays of 
$\Upsilon(10.88)$
show an enhanced affinity for $\Upsilon \pi\pi$, in marked contrast to the decays of other 
$\Upsilon^*$ initial states.
As in the \ssbar and \cc~ cases, the mass of 10.88GeV agrees with the hybrid prediction in 
eq(\ref{masses}).

A problem in sorting this out may be that any resonances that happen by chance to lie near the S-wave 
thresholds
will feed, via rearrangement, the  $\pi\pi$ + $\phi; \psi; \Upsilon$ channels. To determine whether 
the source is hybrid or not
will require measuring, or placing limits on, analogous rearrangement channels to spin-0 onia states 
and comparison
to their spin-1 analogues. Thus for the \bb~ case, the appearance 
of a signal in $\Upsilon \pi\pi$ suggests that a spin-1 \bb~ initial state rather than a hybrid 
resonance may be driving the phenomenon.

\subsection*{If this 4260 state is not hybrid vector charmonium, then where is it?}

Suppose that it is. Where else should we look? Clearly the $[0,1]^{-+}$ states predicted to lie below
the $Y(4260)$ become interesting. The properties and search pattern for such states are discussed in 
ref.\cite{cgodfrey}.
In $e^+e^- \to \psi + X$ it is possible that such states could feed the signal at 3940MeV. If the
production is via strong flux-tube breaking there is a selection rule\cite{cburns} that suppresses
$\psi + X$ when $X$ has negative parity. However, it is possible that the dominant production for \cc~ 
+ \cc~
is by ``preformation"\cite{bct}, where a perturbative gluon creates the second \cc~ pair (the highly 
virtual photon
having created the initial pair). In such a case there is no selection rule forbidding $X \equiv 
[0,1]^{-+}$ hybrids;
however, the amplitude will be proportional to the short distance wavefucntion of the hybrid, which is 
expected to be small
compared to those of e.g. $\eta_c(3S)$ though perhaps comparable to those of $\chi_J$. 
Thus it would be interesting to measure the
$J^{PC}$ of the $X(3940)$ region to see if it contains exotic $1^{-+}$. Note that both
hybrid mesons and also $\pi$-exchange between flavoured mesons lead us to expect signals in the exotic
$1^{-+}$ channel. Model predictions for production rates are required, or practical ways of
looking for manifestly flavoured (such as double charm or double strange) states in the latter case, 
in order to resolve this conundrum.

\subsection*{What should we do next?}

There are clearly tantalising signals in each of the \ssbar~, \cc~ and \bb~ sectors. The latter
appears, on heavy flavour arguments, to be more likely associated with a conventional S=1 \bb~
resonance than a hybrid. If one could determine the polarization of the outgoing $\Upsilon$
in the $\Upsilon \pi\pi$ final state and compare with the polarization in the $\psi \pi\pi$ and
$\phi\pi\pi$, that could have some strategic interest and stimulate model predictions.

The results of ref\cite{bct} show how the relative decay 
amplitudes to $DD_1, D^*D_{0,1,2}$  may be used to determine the
structure of \cc~ states that are near to the S-wave thresholds. In particular this applies to 
$Y(4260)$ and $Y(4325)$. There are characteristic zeroes that may occur
for vector meson decays:
\bea
&\Gamma(^3S_1 \to D_{1H}D) &= 0\label{firstreln}\\
&\Gamma( ^3D_1 \to D_{1L}D) &= 0\\
&\Gamma(^1\Pi P_1\textrm{(hybrid)}] \to D_1(^1P_1)D) &= 0
\label{zeroes}
\eea
 The first pair of zeroes arise from the affinity of light and heavy $D_{1L},D_{1H}$ for $S$ and 
$D$ couplings
respectively, and the zero in eq.(\ref{firstreln}) was noted by ref. \cite{bgs}. 
For the hybrid decay the result follows from the conclusion of lattice
QCD that decays are driven by \qq~ creation in spin-triplet, which 
implies that a pair of spin-singlets (such as $D$ and $^1 P_1$) cannot be produced from
a spin-singlet, such as a hybrid vector \cc~ . In practice these predictions will be
affected by mixing, which can be determined from other processes
(e.g. see \cite{cs06}), and by phase space. The relative rates are
insensitive to form factor effects at low momenta (see for example refs \cite{ley,bcps,cs06}). 

Hence, if the axial mixing angles are known from elsewhere, the pattern of charm pair production can 
identify the
nature of the decaying $\psi$ state. Determining whether the
\cc~ content of these states is $S=0$ (as for a hybrid) or $S=1$ then follows from the relative
production rates of various combinations of charmed mesons, in particular of their $DD_1$ 
branching ratios.


Given that gluonic (hybrid) states are so confidently predicted to occur in the region of
$\sim 1.5$GeV above the lowest vector meson, I would conclude that such states would
naturally be attracted towards these S-wave thresholds, given their affinity for coupling to 
these very modes\cite{cp95,ikp,ipaton}. The search for hybrids and arguments over interpretation would
have analogues with the competing dynamics in the scalar mesons, $f_0/a_0(980)$, which are
associated with the S-wave $K\bar{K}$ threshold, and the
$D_s(2317)$ and $D_s(2460)$ which appear at the $DK$ and $D^*K$ thresholds. In all of these cases the 
consensus appears
to be that there is a short range QCD ``seed" (be it tetraquarks in the case of the scalars or 
$c\bar{s}$ for the
$D_s$ states) which becomes modified by the coupling to S-wave meson pairs\cite{ctorn}.
I suspect that the vector meson signals that I have discussed here are analogously caused by the 
hybrid seed
coupling with the S-wave mesons near to threshold. However, if it should turn out that they
are not driven by hybrids, then the question of where hybrids are, and how ever they are to be 
isolated, 
will demand serious attention.

\begin{acknowledgments}

I am indebted to the organisers for inviting me to give this talk, and to Chris Thomas and Qiang Zhao 
for discussions.
This work is supported,
in part, by grants from
the UK Science and Technology Facilities Council (STFC), and the
EU-RTN programme contract No. MRTN-CT-2006-035 482: ``Flavianet''. 

\end{acknowledgments}

\bigskip 

\end{document}